\documentclass[%
 reprint,
superscriptaddress,
 amsmath,amssymb,
 aps,
]{revtex4-2}

\usepackage{xcolor}
\usepackage{graphicx}% Include figure files
\usepackage{dcolumn}% Align table columns on decimal point
\usepackage{bm}% bold math
\usepackage{MnSymbol}
\begin{document}

\preprint{APS/123-QED}

\title{Viscous-inertial transition in dense granular suspension}% Force line breaks with \\

\author{Franco Tapia}
%\altaffiliation[Also at]{Physics Department, XYZ University.}
\affiliation{Earthquake Research Institute, The University of Tokyo, 1-1-1, Yayoi, Bunkyo-ku, Tokyo 113-0032, Japan}
\affiliation{Department of Mechanical Systems Engineering, Tokyo University of Agriculture and Technology, Koganei, Tokyo, Japan}
\affiliation{Aix-Marseille Universit\'e, CNRS, IUSTI, Marseille, France}
\author{Mie Ichihara}
\affiliation{Earthquake Research Institute, The University of Tokyo, 1-1-1, Yayoi, Bunkyo-ku, Tokyo 113-0032, Japan}
\author{Olivier Pouliquen}
\affiliation{Aix-Marseille Universit\'e, CNRS, IUSTI, Marseille, France}
\author{\'Elisabeth Guazzelli} 
\affiliation{Aix-Marseille Universit\'e, CNRS, IUSTI, Marseille, France}
\affiliation{Universit\'e de Paris, CNRS, Mati\`ere et Syst\`emes Complexes (MSC) UMR 7057, Paris, France}

\date{\today}% It is always \today, today,

\begin{abstract}
Granular suspensions present a transition from a Newtonian rheology in the Stokes limit to a Bagnoldian rheology when inertia is increased. 
A custom rheometer which can be run in a pressure- or a volume-imposed mode is used to examine this transition in the dense regime close to jamming. By varying systematically the interstitial fluid, shear rate, and packing fraction in volume-imposed measurements, we show that the transition takes place at a Stokes number of 10 independent of the packing fraction. Using pressure-imposed rheometry, we investigate whether the inertial and viscous regimes can be unified as a function of a single dimensionless number based on stress additivity.
%Dense granular suspensions can exhibit different regimes depending on the boundary conditions and stress distribution. In general, the flow is mainly controlled by the ratio of the shear rate and the particle pressure, which could be partially well described by a frictional approach for a dilatant granular media. However, as the shear rate increases and the fluid viscosity decreases, the flow can transit from a viscous to an inertial regime. In absence of constitutive equations, dimensional analysis does not give the crucial information to determine this transition, and except for some numerical works, there is not substantial experimental evidence of this phenomenon. 
%In the present work, we present an experimental evidence of the viscous-inertial transition for suspension of non-colloidal rigid particles. A specially  designed pressure-and-volume imposed rheometer is used to explore the dense regime. By varying systematically the interstitial fluid,  shear rate and packing fraction, we show that the transition takes place at a specific Stokes number, which is independent of the packing fraction. The algebraic power law for the viscosity divergence is also shown to be independent on the regime. 
\end{abstract}

\maketitle

Dense granular suspensions that consist of concentrated mixtures of non-colloidal particles suspended in a liquid are ubiquitous in many natural phenomenon (landslides, debris flows, sediment transport) and industrial processes (concrete, pastes). Their rheology is not fully understood and establishing a unified theoretical framework across the different flowing regimes is still challenging. In the viscous regime \cite{GuazzelliPouliquen2018}, the rheology is Newtonian as the (shear and normal) stresses scale linearly with the shear rate, $\dot{\gamma}$, at a constant volume fraction, $\phi$, i.\,e.\,exhibiting a viscous scaling $\propto \eta_f  \dot{\gamma}$ where $\eta_f$ is the viscosity of the suspending fluid and with the prefactor being the relative viscosity which is a sole function of $\phi$. In the inertial regime for which the exemplary case is a dry granular flow \cite{ForterrePouliquen2008}, the rheology is Bagnoldian \cite{Bagnold1954} as the stresses scale quadratically with $\dot{\gamma}$, and more precisely vary as $\propto \rho_p d^2 \dot{\gamma}^2$ where $\rho_p$ is the particle density and $d$ their diameter.

The transition between these two regimes where there are a great number of environmental and engineering applications is not well deciphered. It is considered to take place when the viscous and inertial stresses have the same order of magnitude, i.\,e.\,to be described by the ratio between the inertial and viscous stress scales, namely the Stokes number $St=\rho_p d^2 \dot{\gamma}/\eta_f$. However, since the seminal work of Bagnold \cite{Bagnold1954} who identified the two regimes, the studies are scarce and still inconclusive. Numerical studies \cite{Trulssonetal2012,Amarsidetal2017,NessSun2015} finds that it happens at a transitional Stokes number $St_{v \rightarrow i} \sim 1-2$ independent of $\phi$, while two experiments \cite{Falletal2010,Madrakietal2020} and theoretical work for frictionless particles \cite{DeGiulietal2015}  suggest that $St_{v \rightarrow i}$ vanishes when approaching the jamming transition where $\phi \rightarrow \phi_c$.

In this work, we perform volume-imposed rheometry of dense granular suspensions having suspending fluids of variable viscosity to show unambiguously that the transition from Newtonian to Bagnoldian rheology occurs at $St_{v \rightarrow i} = 10$ independent of $\phi$. We then  conduct pressure-imposed rheometry \cite{BoyeretalPRL2011,Tapiaetal2019} to examine whether the inertial and viscous regimes can be unified as a function of a single dimensionless number based on stress additivity as suggested by numerical simulations \cite{Trulssonetal2012,Amarsidetal2017,Voetal2020}.

\begin{figure}
 \includegraphics[width=0.45\textwidth]{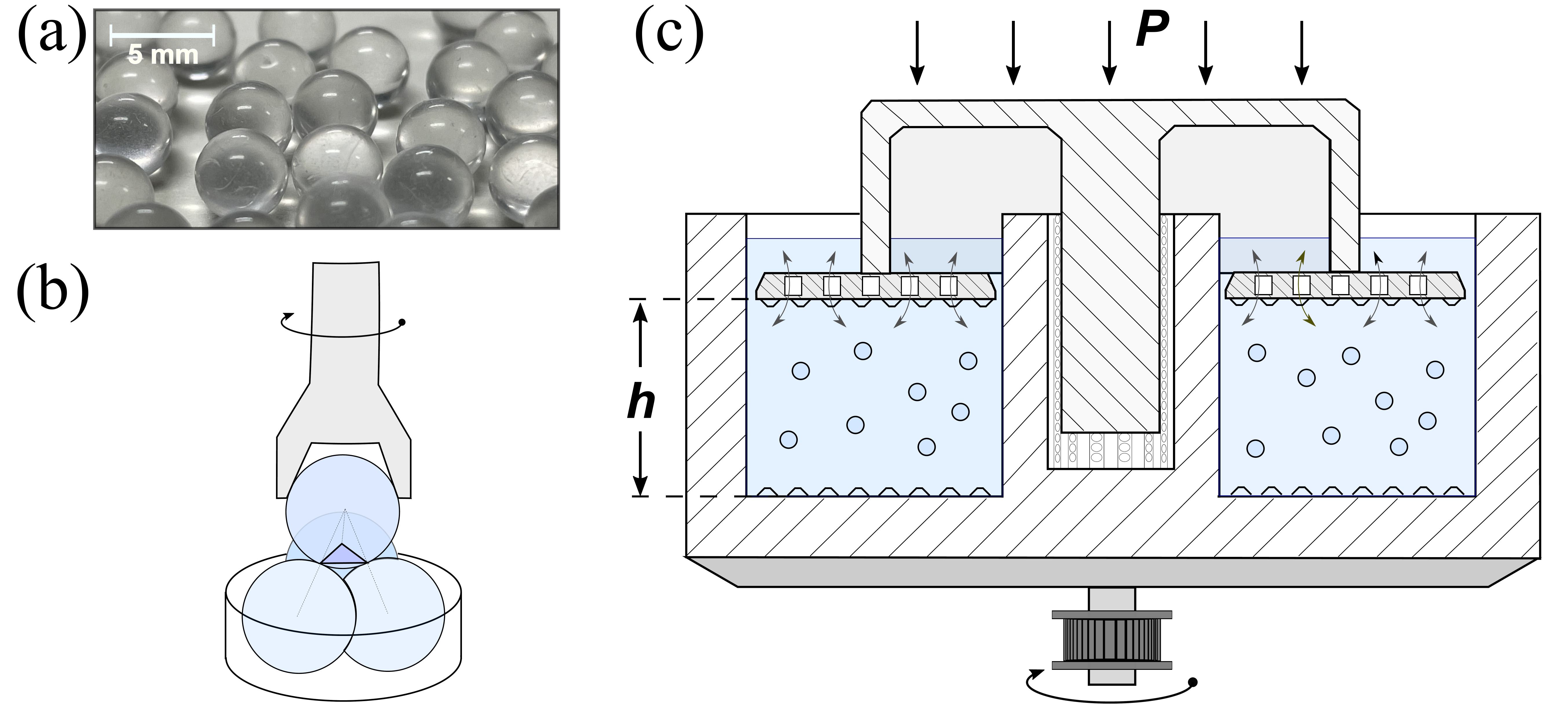}
  \caption{\label{fig:exp} (a) Image of the Poly(methyl methacrylate) spheres. (b) Sketch of the four-ball tribometer. %in the configuration of an equilateral tetrahedron where the upper sphere rotates while staying in contact with lower fixed packing of three spheres 
(c) Sketch of the rheometer. }
\end{figure}

The suspensions used in the experiments consisted of large Poly(methyl methacrylate) spheres, shown in Fig.\,\ref{fig:exp}(a), having mean diameter $d=4.65$\,mm (with a dispersion in size of $1\%$)  and density $\rho_p=1181$\,kg/m$^3$ dispersed in a Newtonian mixture of water, UCON oil, and some amount of sugar insuring density matching with the particles. Four different fluids were used with varying viscosity, $\eta_f=11,  20, 50,$ and $86$\,mPa\,s. Additional negatively-buoyant systems involving the sole dry spheres ($\eta_f=0.0183$\,mPa\,s) and the spheres immersed in pure water ($\eta_f=1$\,mPa\,s) were utilized in pressure-imposed rheometry. The properties of these particles were characterized as done in \cite{Tapiaetal2019}, see the Supplemental Material for details.  We quantified the particle surface shape with confocal scanning microscopy and found surface roughnesses characterized by an average roughness $R_a=0.29\,\mu$m and a standard deviation $R_q=0.51\,\mu$m. We performed controlled sliding experiments using a four-ball tribological tester monitored by a stress-controlled rheometer (AR2000ex, TA Instruments), see Fig.\,\ref{fig:exp}(b), and measured a similar sliding-friction coefficient $\mu_{sf} =0.47\pm0.06$ in dry and water-immersed conditions but a smaller $\mu_{sf} =0.21\pm0.04$ with UCON mixtures.

\begin{figure*}
\includegraphics[width=0.9\textwidth]{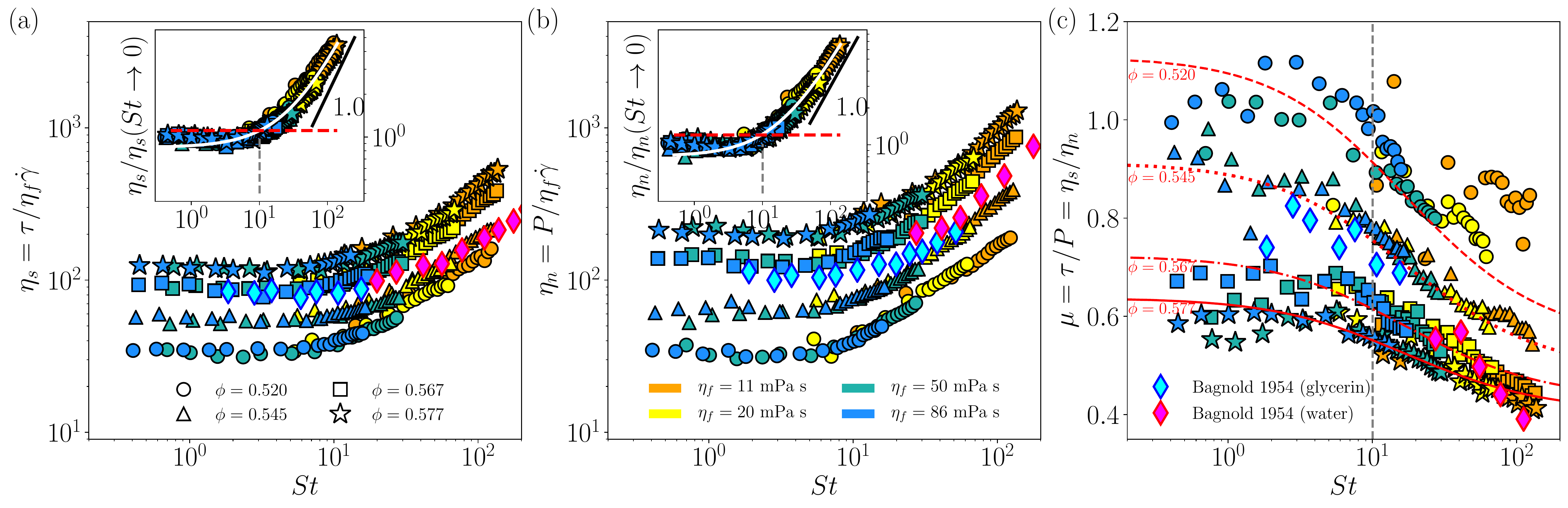}
\caption{\label{fig:phi-imposed} Rheological data coming from $\phi$-imposed rheometry: (a) $\eta_s=\tau/\eta_f \dot{\gamma}$, (b) $\eta_n=P/\eta_f \dot{\gamma}$, and (c) $\mu=\tau/P= \eta_s/\eta_n$ versus $St=\rho_p d^2 \dot{\gamma}/\eta_f$ for 4 suspensions having suspending fluids of variable viscosity, $\eta_f=11\,\mbox{(orange)},  20\,\mbox{(yellow)}, 50\,\mbox{(green)}, 86\,\mbox{(blue)}$\,mPa\,s, at 4 different volume fractions, $\phi=0.520\,(\bigcirc), 0.545\,(\triangle), 0.567\,(\square), 0.577\,(\largestar)$. The data ($\lozenge$) from Fig.7 of \cite{Bagnold1954} for spheres having $d=1.32$\,mm and two suspending fluids $\eta_f=1$ (magenta) and $7$ (cyan)\,mPa\,s are also reported. Insets of graphs (a,b): $\eta_s/\eta_s (St\rightarrow0)$ and $\eta_n/\eta_n (St\rightarrow0)$  versus $St$ showing a transition from a Newtonian (slope 0) to a Bagnoldian (slope 1) rheology. Red lines in (c): Constitutive law (\ref{ eq:CLmu}) for the different $\phi$ with $\phi_c=0.615,\mu_c=0.31, a_{\mu}=11.29,a_{\phi}=0.66,\alpha_{\mu} =0.0088,\alpha_{\phi}=0.1$ coming from the $P$-imposed rheology.}
\end{figure*}

The rheological measurements were undertaken with a custom rheometer \cite{Tapiaetal2019} depicted in Fig.\,\ref{fig:exp}(c). This shearing device is composed of an annular cylinder (of inner radius 43.95 mm and outer radius 90.28 mm) that is attached to a bottom plate and covered by a top permeable plate. Both plates are made rough by using wire meshes with openings of 6.3 mm corresponding to 1.3 $d$. The top plate can be moved vertically with a linear positioning stage and enables the fluid to flow through it but not the particles.  The bottom plate can be rotated at a constant angular velocity to produce a linear shear. The range of shear rate that can be achieved is $1 \leqslant \dot{\gamma} \leqslant 50$\,s$^{-1}$. The thickness of the cell, $h$, can be varied between 24.8 and 27.5 mm, i.\,e. $5.3\,d \leqslant h \leqslant 6\,d$. The shear stress, $\tau$, is deduced from measurements of the torque exerted on the top plate after calibration with the pure fluid. The component of the normal stress perpendicular to the top plate, simply referred as the particle pressure $P$, is given by a precision scale attached to the translation stage after correction for buoyancy. The volume fraction, $\phi$, can be deduced from the position of the top plate recorded by a sensor. We estimated the accuracy based on the fluctuations around the stationary values to be 7\%, 3\%, and $\pm0.002$ on the measurements of $\tau$, $P$, and $\phi$, respectively. The rheometer could be run in a $P$-imposed or a $\phi$-imposed mode using a feedback control loop involving the scale measurement or the position of the top plate, respectively. It was operated in an air-conditioned room at 24$^{\circ}$\,C and evaporation of the suspending fluid was inhibited by using a solvent trap covering the cell. The rheological data, data analysis, and plots are given in the Supplemental Material.

We start by presenting in Fig.\,\ref{fig:phi-imposed}(a,b) the rheological data collected in $\phi$-imposed rheometry for the suspensions having suspending fluids of variable $\eta_f$ at different $\phi$. A Newtonian regime in which the shear and normal viscosities, $\eta_s=\tau/\eta_f \dot{\gamma}$ and $\eta_n=P/\eta_f \dot{\gamma}$ respectively, are independent of $St$ (i.\,e.\,independent of $\dot{\gamma}$) but increase with increasing $\phi$ is observed for $St < 10$. Around $St \approx 10$, the rheology transitions smoothly from Newtonian to continuously shear thickening. For $St > 10$, a Bagnoldian regime in which $\eta_s$ and $\eta_n$ scale linearly with $St$ (i.\,e.\,with $\dot{\gamma}$) is reached.  We have also plotted in Fig.\,\ref{fig:phi-imposed}(c) the effective friction coefficient $\mu=\tau/P= \eta_s/\eta_n$ versus $St$. We recover the Stokes regime for $St < 10$. For larger $St$, $\mu$ decreases with increasing $St$ as $\eta_s$ presents a slower transition to the Bagnoldian regime than $\eta_n$.

We have also reported in these graphs the data coming from Fig.7 of Bagnold's paper \cite{Bagnold1954} obtained in the the transition region for a single $\phi=0.555$ ($\lozenge$ magenta and cyan). Strikingly, although they were obtained with different particles (spheres of nearly $50\%$ mixture of paraffin wax and lead stearate with $d=1.32$ mm and $\rho_p$ close to that of water), different fluids (water with $\eta_f=1$\,mPa\,s and glycerin mixture with $\eta_f=7$\,mPa s), and a different experimental device, these data show a similar transition around $St \approx 10$ and are located in between the present data at $\phi=0.545$ and $0.567$. 

\begin{figure*}
\includegraphics[width=0.90\textwidth]{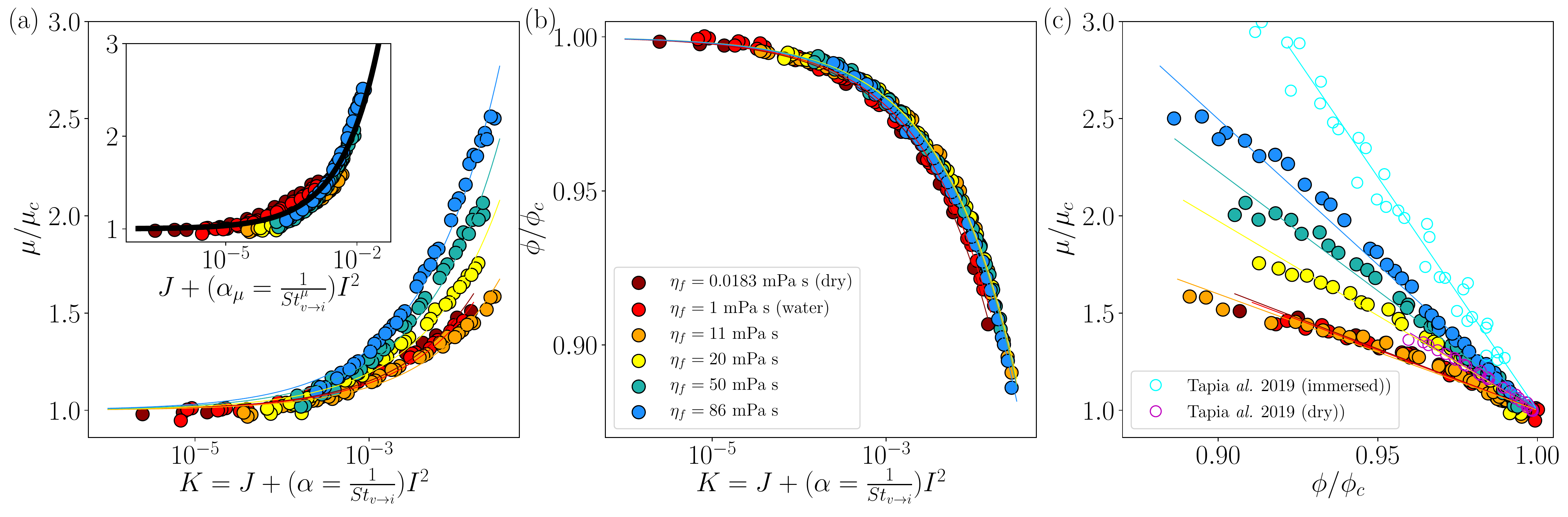}
\caption{\label{fig:P-imposed} Rheological data coming from $P$-imposed rheometry: (a) $\mu/\mu_c$ and (b) $\phi/\phi_c$ versus $K=J+\alpha I^2$ with $\alpha=1/St_{v \rightarrow i}$ and (c) $\mu/\mu_c$ versus $\phi/\phi_c$ for the same neutrally-buoyant suspensions as in Fig.\,\ref{fig:phi-imposed} and two additional negatively-buoyant systems consisting of the sole dry spheres and of the spheres immersed in pure water. The data coming from Fig.5 of \cite{Tapiaetal2019} using slightly rough polystyrene sphere (with $d=580\,\mu$m) in the dry case as well as immersed in a very viscous Newtonian fluid of same density (with $\eta_f=2.01$ Pa s) are also reported. Inset of (a): $\mu/\mu_c$ versus $J+\alpha_{\mu} I^2$; the black solid line corresponds to the best fit: $\mu/\mu_c=1+ a_{\mu} (J+\alpha_{\mu}I^2)^{1/2}$ with $a_{\mu}=11.29$ and  $\alpha_{\mu}=0.0088$. }
\end{figure*}

The fact the transition takes place at a specific Stokes number independent of $\phi$ is evidenced in the insets of Fig.\,\ref{fig:phi-imposed}(a,b) where the viscosities normalized by their values when $St\rightarrow0$ collapse onto master curves exhibiting the transition from a Newtonian (slope 0) to a Bagnoldian (slope 1) rheology. To find accurately the transitional $St_{v \rightarrow i}$, we choose a threshold value of 3 standard deviations above the mean for the low-$St$ data and compute the point of interception of the horizontal line at this value (horizontal red dashed line) with the simplest power-law fit of the whole data (white solid line). We find $St_{v \rightarrow i}=10.0$ with an accuracy $\pm0.2$ and $\pm0.5$ for $\eta_s(St)$ and $\eta_n(St)$, respectively.

\begin{table}
\caption{\label{tab:coeff} Coefficients of the linear regressions in $K^{1/2}$.}
\begin{ruledtabular}
\begin{tabular}{ccccc}
&\multicolumn{2}{c}{$\mu = \mu_c(1 + a_{\mu}^K K^{1/2})$}&\multicolumn{2}{c}{$\phi=\phi_c(1-a^K_{\phi} K^{1/2})$ }\\
 $\eta_f$ (\,mPa\,s)&$\mu_c$&$a_{\mu}^K$&$\phi_c$&$a^K_{\phi}$\\ \hline
 0.0183 & 0.39 & 4.76  &0.596&0.75\\
1& 0.41 &4.41 & 0.586 &0.71\\
11& 0.31& 3.79 & 0.615 & 0.63\\
20& 0.31 & 6.07 & 0.618&0.62\\
50& 0.31 &  7.85 & 0.615&0.64\\
86& 0.31 & 9.96 &0.614 &0.66\\
\end{tabular}
\end{ruledtabular}
\end{table}

We now turn to the examination of rheological measurements obtained in $P$-imposed rheometry.  When sheared under a confining pressure $P$, the rheological response of a particulate system is determined by two dimensionless quantities: the effective friction coefficient, $\mu=\tau/P$, and the packing fraction, $\phi$. When inertia dominates as for a dry granular flow, these quantities are expected to depend solely upon $I^2= \rho_p d^2 \dot{\gamma}^2/P$ which is the ratio of the inertial stress scale $\sim \rho_p d^2 \dot{\gamma}^2$ and the external confinement pressure $\sim P$ \cite{ForterrePouliquen2008}. When viscous effects prevail, a viscous stress scale $\sim \eta_f  \dot{\gamma}$ should be used instead, and the control parameter is $J= \eta_f  \dot{\gamma}/P$ \cite{BoyeretalPRL2011}. It has been conjectured \cite{Trulssonetal2012,Amarsidetal2017,Voetal2020} that these two regimes can be unified by using a single dimensionless parameter based on stress additivity, $K=J+\alpha I^2$, and that the transition from viscous to inertial flow happens at a Stokes number $I^2/J=1/ \alpha=St_{v \rightarrow i}$.

In Fig.\,\ref{fig:P-imposed}(a,b), we show the $P$-imposed rheological data for $\mu$ and $\phi$ against $K=J+\alpha I^2$ with $\alpha=1/St_{v \rightarrow i}=0.1$ for the same neutrally-buoyant suspensions as in Fig.\,\ref{fig:phi-imposed} and two additional negatively-buoyant systems consisting of the sole dry spheres and of the spheres immersed in pure water. The data have been normalized by their critical values at jamming, i.\,e.\,by the critical friction coefficient, $\mu_c$,  and the critical (or maximum flowable) volume fraction, $\phi_c$, respectively. These quasistatic values, $\mu_c$ and $\phi_c$, have been obtained by fitting the data using a linear regression in $K^{1/2}$ (shown by the solid lines) as summarized in Table\,\ref{tab:coeff}; since $\mu_c$ and $\phi_c$ depend on the nature of the frictional contact between the particles, $\mu_c$ is smaller and $\phi_c$ larger in the case of frictional contacts lubricated with UCON mixtures with smaller $\mu_{sf}$. While $\phi/\phi_c$ can be expressed as a collapsed function of $K$ given by $\phi/\phi_c=1-a_{\phi} K^{1/2}$ with $a_{\phi}=0.66$, there are different curves for $\mu/\mu_c$ showing a shift in increasing magnitude with increasing $\eta_f$. We have attempted to seek a better collapse by having a different value for $\alpha$. This is obtained by the fit $\mu/\mu_c=1+ a_{\mu} (J+\alpha_{\mu}I^2)^{1/2}$ with $a_{\mu}=11.29$ and  $\alpha_{\mu}=0.0088$, i.\,e. a larger transitional Stokes number $St^{\mu}_{v \rightarrow i}\approx114$, as shown in the inset of  Fig.\,\ref{fig:P-imposed}(a). 

The inability to collapse the data for $\mu$ with $\alpha=1/St_{v \rightarrow i}$ is even more evidenced in Fig.\,\ref{fig:P-imposed}(c) where $\mu/\mu_c$ is plotted against $\phi/\phi_c$ and where we have added the data coming from Fig.5 of \cite{Tapiaetal2019} using slightly rough polystyrene spheres ($d=580\,\mu$m and $\rho_p=1050$\,kg/m$^3$) in the dry case as well as immersed in a very viscous Newtonian fluid of same density ($\eta_f=2.01$ Pa s). The data align in straight lines jointed at the jamming point which present an increasing slope with increasing $\eta_f$ between the two bounds given by the lowest slope for the dry case and the highest slope for the most viscous case.

\begin{figure}
\includegraphics[width=0.3\textwidth]{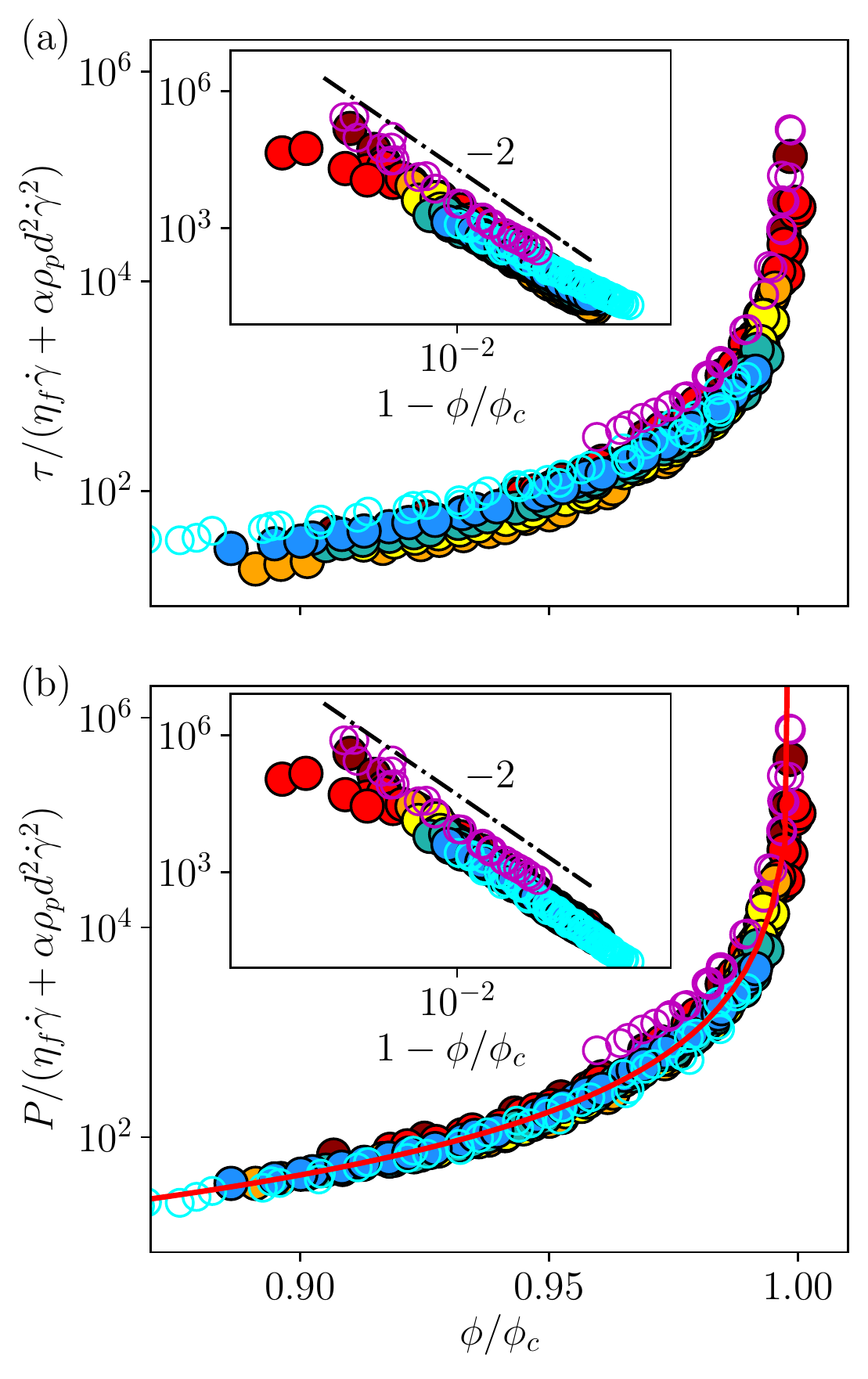}
\caption{\label{fig:P-imposed-viscosities} Rheological data coming from $P$-imposed rheometry: (a) $\tau$ and (b) $P$ normalized by $\eta_f \dot{\gamma} + \alpha \rho_p d^2 \dot{\gamma}^2=\eta_f \dot{\gamma} ( 1+ \alpha St)$  versus $\phi/\phi_c$ (same symbols as in Fig.\,\ref{fig:P-imposed}). The insets show these quantities versus the rescaled volume fraction, $1-\phi/\phi_c$. Red solid line in (b): Constitutive law (\ref{eq:CLtauP}) with $a_{\phi}=0.66$.}
\end{figure}

Although the data of Fig.\,\ref{fig:P-imposed} were obtained under $P$-imposed conditions, $\tau$ and $P$ can be inferred and their asymptotic behavior close to the jamming transition examined. Fig.\,\ref{fig:P-imposed-viscosities} shows $\tau$ and $P$ normalized by the addition of stress scales, $\eta_f \dot{\gamma} + \alpha \rho_p d^2 \dot{\gamma}^2$, versus $\phi/\phi_c$. These quantities would represent the shear and normal viscosities, $\eta_s(\phi)=\tau/\eta_f \dot{\gamma}$ and $\eta_n(\phi)=P/\eta_f \dot{\gamma}$, in the purely viscous Newtonian regime and $\phi$-dependent functions, $\eta_I(\phi)=\tau/\rho_p d^2 \dot{\gamma}^2$ and $\eta_{II}(\phi)=P/\rho_p d^2 \dot{\gamma}^2$, in the purely Bagnoldian regime \cite{Tapiaetal2019}. The collapse is excellent for $P$ but poorer for $\tau$. This is due to the imperfect collapse of $\mu$, seen in Fig.\,\ref{fig:P-imposed}(c), since $\tau=\mu P$. Note that this gives only small shifts between the different curves in the $\tau$-graph as the values of $\mu$ are orders of magnitude smaller than those of $\tau$ and $P$. The important output shown in the insets is that the normalized $\tau$ and $P$ functions diverge as $(1-\phi/\phi_c)^{-2}$. The fact that a same divergence is observed across the viscous and inertial regimes shows that the crossover shear-rate, $\dot{\gamma}_{v \rightarrow i} \sim (\eta_f/\rho_pd^2)\, \eta_s(\phi)/\eta_I(\phi)$, or similarly $\sim (\eta_f/\rho_pd^2)\, \eta_n(\phi)/\eta_{II}(\phi)$, is independent of $\phi$ close to the jamming transition. This confirms the finding of a crossover Stokes number independent of $\phi$ obtained in $\phi$-imposed rheometry in Fig.\,\ref{fig:phi-imposed}(a,b).

In this work, we have explored the viscous-inertial transition in dense granular suspension. The first major output coming from $\phi$-imposed rheology is the finding of a crossover from viscous to inertial flow at a Stokes number independent of the packing fraction, $St_{v \rightarrow i}=10$. This qualitatively agrees with numerical simulations of frictional spheres \cite{Trulssonetal2012,Amarsidetal2017,NessSun2015} finding a slightly smaller $St_{v \rightarrow i}\sim1-2$. This is consistent with the data of Bagnold for suspensions having similar large spheres \cite{Bagnold1954} but strongly differs from the experimental results of \cite{Falletal2010,Madrakietal2020} which report decade smaller $St_{v \rightarrow i}\sim10^{-3}-10^{-2}$ for suspensions consisting of much smaller spheres which are potentially non-frictional and prone to colloidal interactions. The second important result coming from $P$-imposed rheology has been to find that the packing fraction is governed by the single dimensionless number $J+\alpha_{\phi} I^2$ with $\alpha_{\phi}=\alpha=1/St_{v \rightarrow i}=0.1$ but not the effective friction which seems to be governed instead by $J+\alpha_{\mu} I^2$ with a smaller $\alpha_{\mu}=0.0088$ characteristic of a larger $St^{\mu}_{v \rightarrow i}\approx114$. The existence of two distinct scalings for $\phi$ and $\mu$ is not captured by the numerical simulations for frictional particles \cite{Trulssonetal2012,Amarsidetal2017,NessSun2015}.

We can then attempt to rationalize the data by proposing the constitutive laws in the dense regime close to jamming,
\begin{eqnarray}
\mu & = & \mu_c [1+ a_{\mu} (J+\alpha_{\mu} I^2)^{1/2}],\\
\phi & = & \phi_c [1- a_{\phi} (J+\alpha_{\phi} I^2)^{1/2}],
\label{eq:CLmuphi}
\end{eqnarray}
for the $P$-imposed rheology.
These relations lead to
\begin{equation}
\mu =\mu_c [1+\frac{a_{\mu}}{a_{\phi}}(1-\frac{\phi}{\phi_c}) \sqrt{(1+\alpha_{\mu} St)/(1+\alpha_{\phi} St)}],
\label{ eq:CLmu}
\end{equation}
which agrees well with the $\phi$-imposed data of Fig.\,\ref{fig:phi-imposed}(c), even though the range of $\phi$ is lower than that of the $P$-imposed data. The shear and normal stresses are then 
\begin{eqnarray}
\tau & = & \mu(\phi, St) \, P(\phi, St),\\
P & = &   \eta_f \dot{\gamma} ( 1+ \alpha_{\phi} St) \, a_{\phi}^2 \, (1-\phi/\phi_c)^{-2}, 
\label{eq:CLtauP}
\end{eqnarray}
in agreement with the scaling and asymptotic behavior seen in Fig.\,\ref{fig:P-imposed-viscosities}. The slower transition of $\tau$ compared to that of $P$ is embedded in $\mu (\phi, St)$ which describes the anisotropy of the stresses and may reflect changes in the suspension microstructure and in the nature of the particle interactions when increasing inertia.

\vspace{0.5cm}
We greatly thank M. Kameda and O. Kuwano for discussions and W. Lecoz and S. No\"el for technical assistance. This work was supported by the Labex MEC (ANR-10-LABX-0092) under the A*MIDEX project (ANR-11-IDEX-0001-02) funded by the French government `Investissements d'Avenir programme'  and by the JSPS KAKENHI Grant Number 19H00713. FT is supported by the visiting researcher program of the Earthquake Research Institute.

\bibliography{PRLTapiabib}
\end{document}